\documentclass{optica-article}
\usepackage{amsmath,bm}

\journal{opticajournal}
\articletype{Research Article}
\usepackage{lineno}
\begin{document}

\title{Inverse design of optical pulse shapes for time-varying photonics}

\author{Joshua Baxter,\authormark{1,*} and Lora Ramunno,\authormark{1,**}}

\address{\authormark{1}Department of Physics, and Nexus for Quantum Technologies Institute, University of Ottawa, Ottawa, Canada}

\email{\authormark{*}jbaxt089@uottawa.ca \\ \authormark{**}Lora.Ramunno@uottawa.ca} %% email address is required; see note below about the corresponding author designation



\begin{abstract}
There has been an explosion of interest in time-varying photonics due to the recent discovery and design of materials and metamaterials with strong, time-varying, nonlinear optical responses. This opens the door to novel optical phenomena including reciprocity breaking, frequency translation, and amplification that can be enhanced by optimizing the light-matter interaction. Although there has been recent interest in applying topology-based inverse design to this problem, we have decided to take a more novel approach.  In this article, we will introduce a method for the inverse design of optical pulse shapes to enhance their interaction with time-varying media. We test our objective-first approach by maximizing the transmittance of optical pulses of equal intensity through time-varying media. Indeed, without requiring a change in pulse energy, we demonstrate large, broadband enhancements in the pulse energy transmission through the thin-films, including gain. Our final test includes maximizing pulse transmission through indium tin oxide, a time-varying medium when strongly pumped in its ENZ band. Through this work, we hope to inspire exploration of this new degree of freedom.
\end{abstract}

%%%%%%%%%%%%%%%%%%%%%%%%%%  body  %%%%%%%%%%%%%%%%%%%%%%%%%%
\section{Introduction}
Until recently, nanophotonics research has been mainly focused on the spatial confinement of light via plasmonics, dielectric resonators, and waveguides. The space of possible devices is virtually infinite, and novel techniques have been introduced to explore this parameter space in non-intuitive ways. One such method is adjoint sensitivity analysis, which has gained recent interest with the successful design of integrated optical devices including wavelength demultiplexers \cite{piggott_inverse_2015},  demultiplexing grating couplers \cite{piggott_inverse_2014}, y-splitters \cite{lalau-keraly_adjoint_2013}, and
reflectors/resonators \cite{ahn_photonic_2022}. Recently, this method has been extended to nonlinear optical devices whereby the intensity of the incoming light determines the functionality of the device \cite{hughes_adjoint_2018}. As these methods become more accessible via open-source libraries \cite{su_nanophotonic_2020,hammond_high-performance_2022}, textbooks \cite{bakr_adjoint_2017}, and tutorials \cite{christiansen_inverse_2021}, we should expect non-intuitive device design to become the norm in nanophotonics. Adjoint sensitivity analysis has been developed for the design of two- and three-dimensional geometries, however, there is a fourth dimension that has yet to be exploited: time. 
\newline

In recent years, a new branch of photonics has emerged that investigates wave propagation through time-varying media \cite{galiffi_photonics_2022,ferrera_dynamic_2017,shaltout_spatiotemporal_2019,badloe_tunable_2021} driven by developments and promises of 4D metamaterials \cite{engheta_metamaterials_2021,shaltout_time-varying_2015,fan_active_2022}, and recent discoveries of materials with optically-driven, ultrafast permittivity modulations \cite{alam_large_2016,alam_large_2018,ferrera_ultra-fast_2018}. Time-varying materials have been shown to host unique physical effects including frequency translation \cite{zhou_broadband_2020}, reciprocity/time-reversal symmetry breaking \cite{shaltout_time-varying_2015} and have applications in optical isolation \cite{mock_optical_2020}, all-optical switching \cite{bohn_all-optical_2021}, and beam steering \cite{sabri_broadband_2021}, to name a few.
\newline

There is now interest in applying inverse design techniques for the geometric optimization of time-varying media \cite{li_empowering_2022}. This is an exciting prospect, though challenging due to the computational/memory requirements involved. Nonetheless, geometric optimization via time-domain adjoint sensitivity analysis has been explored \cite{bakr_adjoint_2004,bakr_fdtd-based_2016,hassan_topology_2022}, and these techniques will most likely be applied to the geometric optimization of time-varying media in the near-future. That said, geometric optimization is only half of the story when it comes to controlling light - matter interaction; we can also control the light. 
\newline

In static, linear materials, the shape of the pulse is irrelevant to the device output as it can always be decomposed into its Fourier components. In time-varying media, this is no longer the case. As a trivial example, two identical pulses delayed with respect to each-other (a simple form of pulse-shaping) and incident onto a time-varying thin film may experience different refractive indices and thus each pulse will have a different transmittance/reflectance spectrum, despite being spectrally equivalent. As a less trivial example, one may conceive of a pulse chirped so that its central frequency in the time-varying film will follow the transmittance peak. As one could imagine, the light-matter interaction in this case is quite complicated and finding optimal 
pulse shapes thus necessitates computational optimization.
\newline

In this paper, we introduce a new paradigm in inverse design to optimize the shape of light pulses based on a new application of adjoint sensitivity analysis. We present protocols for the inverse design of the shapes of pulses incident on time-varying optical materials that can be used, for example, to enhance or minimize the amount of light that is transmitted through such a time-varying medium. This is done via
two approaches: 
1) optimizing the pulse in time, where the field of the pulse at each (discretized) time-step is tuned, and 2) optimizing the pulse in frequency, where the phase of each (discretized) frequency component is tuned as in 4f pulse shaping. The first approach allows for the full exploration of parameter space in time, presenting us with non-intuitive pulse shapes that are not restricted to the frequency spectrum of the initial pulse shape. The second approach conserves the frequency spectrum of the initial pulse. In Section 2, we present the adjoint sensitivity analysis for time-varying materials, We outline how they are implemented for pulse shaping in time (Section 3) and in frequency (Section 4), where we test both approaches through a series of toy-model examples and analyze the behaviour of the optimized pulse. Finally, in Section 5, we apply these two methods to the realistic example of light propagation through a thin film of indium tin oxide (ITO), a material synonymous with time-varying photonics due to its high refractive index perturbations when excited near its epsilon-near-zero (ENZ) band \cite{alam_large_2016,alam_large_2018}.

\section{Adjoint method for time-varying materials}

To optimize any sort of design, be it spatial or temporal, it is useful to calculate the gradient of the objective function $F$ with respect to the tunable parameters. Once this gradient is known, there are many iterative, gradient-based optimization algorithms that can be used for the design process. Adjoint sensitivity analysis provides a very convenient method for obtaining this gradient with just two simulations, for an arbitrarily large number of tuning parameters, and accordingly has been investigated extensively for topology optimization of static optical systems.
\newline

Since we wish to shape pulses incident on time-varying optical materials, we employ a time-domain adjoint sensitivity analysis, wherein the objective function $F$ is defined as the integral over a time-dependent objective function $\psi$, which in general depends on a set of $M$ tunable parameters $\textbf{p}=[p_0,…,p_i…,p_M]$, as well as the time- and space-varying electric ($\textbf{E}$) and magnetic ($\textbf{H}$) fields, and possibly additional auxiliary fields \cite{hassan_topology_2022}. The gradient of $F$ is then described by \cite{bakr_adjoint_2017}

\begin{equation}\label{generalgrad}
    \frac{dF}{d\textbf{p}} = \Big[\frac{dF}{dp_0},...,\frac{dF}{dp_i},...,\frac{dF}{dp_M}\Big] = \frac{d}{d\textbf{p}}\int_0^{T_m}\psi(\textbf{p},\textbf{E},\textbf{H},..) dt,
\end{equation}

\noindent where $T_m$ is the maximum simulation time, and the fields are assumed to be zero before $t=0$ and after $t=T_m$.  The functional form of $\psi$ is known and set according to what we desire to optimize.
\newline

In the supporting information (SI) Section \ref{td-adjoint} we outline an adjoint sensitivity analysis for two classes of time-varying materials, one that can be described by a “dispersionless” time-varying permittivity $\varepsilon(t)$ (SI Section \ref{2.1}), and another that can be described by time-varying dispersive models for the current density or polarization fields (SI Section \ref{2.2}). For the latter case, we derive explicitly the equations for the Drude model with a time-varying plasma frequency $\omega_p(t)$.
\newline

Our aim is to tune the time signal of the source current $\textbf{J}_s(t)$ in order to maximize some quantity (e.g. the transmittance), rather than tuning the geometric topology. This is done via two simulations: a \textit{forward} simulation that is a direct solution of Maxwell's equations with a current source $\textbf{J}_s(t)$, and an \textit{adjoint} simulation that is a solution to an adapted form of Maxwell's equations with a current source (see Eq. \ref{adjointsource})

\begin{equation}
    \textbf{J}_s^{adj}=-\frac{\partial\psi}{\partial\textbf{E}} (T_m-t)
    \label{adjoint_current_main}
\end{equation}

placed at the spatial location where the objective function is defined. Note that $\textbf{J}_s^{adj}$ is time-reversed for reasons made clear in its derivation.
\newline

For both time-varying permittivity and time-varying dispersion, we have a prescription for calculating the full gradient $\partial F/\partial \textbf{p}$ with only two simulations. The steps involved are  

\begin{enumerate}
    \item compute the forward problem (Eq. \ref{maxwell1} or \ref{maxwell2}) to find the forward fields, and evaluate the external current density source of the adjoint problem (Eqs. \ref{adjoint_current_main} and \ref{adjointsource}),
    \item compute the adjoint problem (Eq. \ref{adjoint1} or \ref{adjoint2}) to find the adjoint fields, with minor modifications made to the simulation software if this is not the same as the forward problem, 
    \item compute the gradient (Eq. \ref{grad} or \ref{grad2}) using the time-reversed adjoint fields and the residue vector calculated from the forward fields (Eq. \ref{residuefull} or \ref{residuefull2}).
\end{enumerate}

Our pulse-shaping inverse design methods are implemented in a one-dimensional (1D) finite - difference time - domain (FDTD) code \cite{computational_nanophotonics_videos_write_2021}. The python scripts for all examples of Sections 3 - 5 are made available in the Supplementary Materials. 
%\newline 

We consider 1D examples for the sake of simplicity in introducing this method, and for simple interpretation of the code we are making available. In fact, 2D and 3D implementations of this method are entirely possible, depending upon the spatial extent of the objective function. However, 1D is also useful in itself because much current work in active nanophotonic % works 
has involved planar geometries like thin-films \cite{alam_large_2016,baxter_dynamic_nodate} and metasurfaces \cite{shaltout_spatiotemporal_2019,alam_large_2018}, which can be described using bulk (for thin-films) or effective \cite{smith_determination_2002,smith_electromagnetic_2005} (for metasurfaces) refractive indices. This enables the problem to be scaled to 1D wherein our provided pulse-shaping code can be directly applied.
\newline

In our code, we take the propagation direction to be along y, and our incident current density $\textbf{J}_s$ to be linearly polarized in the x-direction. We define our instantaneous objective function $\psi(t)$ as the square of the x-component of $\textbf{E}$ at a single location $y_{trans}$ at least a few wavelengths away from the thin film. We then set the objective function as 

\begin{equation}
    F = \frac{1}{2}\int_0^{T_m} E_x^2(y_{trans},t) dt,
\end{equation}

\noindent which is proportional to the transmitted pulse energy, and the factor of 2 is included for future convenience. 
\newline

The source current of the adjoint simulation is thus applied only at the single location $y_{trans}$, and using Eq. \ref{adjointsource} we find it is given by 

\begin{equation}
    J_s^{adj}(t) = -E_x(y_{trans},T_m - t),
\end{equation}

\noindent where $E_x(y_{trans},t)$ is stored from the forward simulation. In order to accomplish pulse shaping, we allow the external source of our forward simulation, $J_s(t)$, to vary with respect to a set of tuning parameters, $p_i$. Assuming $J_s(t)$ is applied at a single point $y_s$, the gradient of the objective function (see Eqs. \ref{grad} or \ref{grad2}) becomes

\begin{equation}\label{grad3}
    \frac{\partial F}{\partial p_i} = -\int_0^{T_m}E_x^{adj}(y_s,T_m-t)\frac{\partial J_s (t)}{\partial p_i} dt.
\end{equation}

In the sections that follow, we introduce two strategies for parameterizing $J_s(t)$. In Section 3 we take the set $p_i$ to be the actual time-domain values of the (normalized) pulse signal at discrete times $t_i$. In Section 4 we take $p_i$ to be a phase factor applied to the $i^{th}$ frequency component of the pulse, such as would be introduced in a 4f pulse shaping setup. 

\section{Pulse Shaping in Time}

In this section, we optimize our objective function by directly tuning the pulse shape as a function of time. We take the values of the pulse at each discretized moment in time $t_i$ as our tuning parameters $p_i$ though other methods are possible (for example, restricting the pulse to a particular functional form with the fitting parameters as the $p_i$). To conserve energy of the pulse throughout the optimization process, however, the pulse must be normalized at each optimization step. Our external current density source is then taken as

\begin{equation}
    J_s(t)=J_0\frac{p(t)}{\bar{p}},
\end{equation}

\noindent where $p(t)$ is the unitless pulse shape that we are optimizing, and $\bar{p}^2 = 1/T_m\int_0^{T_m}p^2(t)dt \approx 1/N_T\sum_{i=0}^{N_T}p_i^2$ is an energy normalization factor, where $N_T$ is the maximum number of time steps in the simulation. Our optimization parameters are then $p_i = p(t_i)$ for discrete values of time $t_i$. The current density amplitude is $J_0$, which is chosen such that $\int J_s^2(t) dt = 1$.
\newline

Asssuming fixed geometrical topology, and a source located at $y_s$ in our 1D forward simulation, the gradient according to Eq. \ref{grad3} becomes

\begin{equation}
    \frac{\partial F}{\partial p_i} = -E_x^{adj}(y_s,T_m-t_i)\frac{J_0}{\bar{p}}\Big[1-\frac{1}{N_T}\frac{p_i^2}{\bar{p}^2}\Big].
\end{equation}

\noindent This expression gives us the sensitivity of the objective function $F$ (proportional to the transmitted pulse energy) with respect to $p_i$ (the incident pulse at each time step). With this gradient we can now tune each $p_i$, and thus the incident pulse shape, in order to maximize the $F$ and thus the transmitted pulse energy.
\newline

Though we have ensured that the incident pulse energy remains constant, we have made no restrictions on its frequency components or bandwidth. In fact, tuning the pulse using this gradient can and will result in the creation of new frequency components. While this is not so useful for optimizing a given pulse source with a given bandwidth (a scenario we return to in Section 4), it could be useful for exploring, for example, what kind of source with what bandwidth would be ideal for a given geometry and material response. It also presents some physically interesting results that we will describe below. 
\newline

In the following, we use our pulse shaping method to maximize the transmitted pulse energy through time-varying dielectric (Section \ref{time_varying_perm1}), and metallic (Section \ref{time_varying_drude1}) thin films. We will restrict ourselves to test-models commonly used in time-varying photonics: sinusoidal varying permittivities \cite{koutserimpas_parametric_2018,pendry_gain_2021,galiffi_photonics_2022}, including with frequency dispersion. In SI Section \ref{staticsection}, for completeness, we demonstrate the effect of pulse shaping when tested on static (time-invariant) thin films where it is found that the pulse changes frequency to match the transmittance resonances of the thin films. In Section 5 we will demonstrate our method on a physical example, ITO pumped by high intensity light in its ENZ spectral region.

\subsection{Test case 1: Time-varying permittivity}\label{time_varying_perm1}
\begin{figure}
\centering
\includegraphics[width = \linewidth]{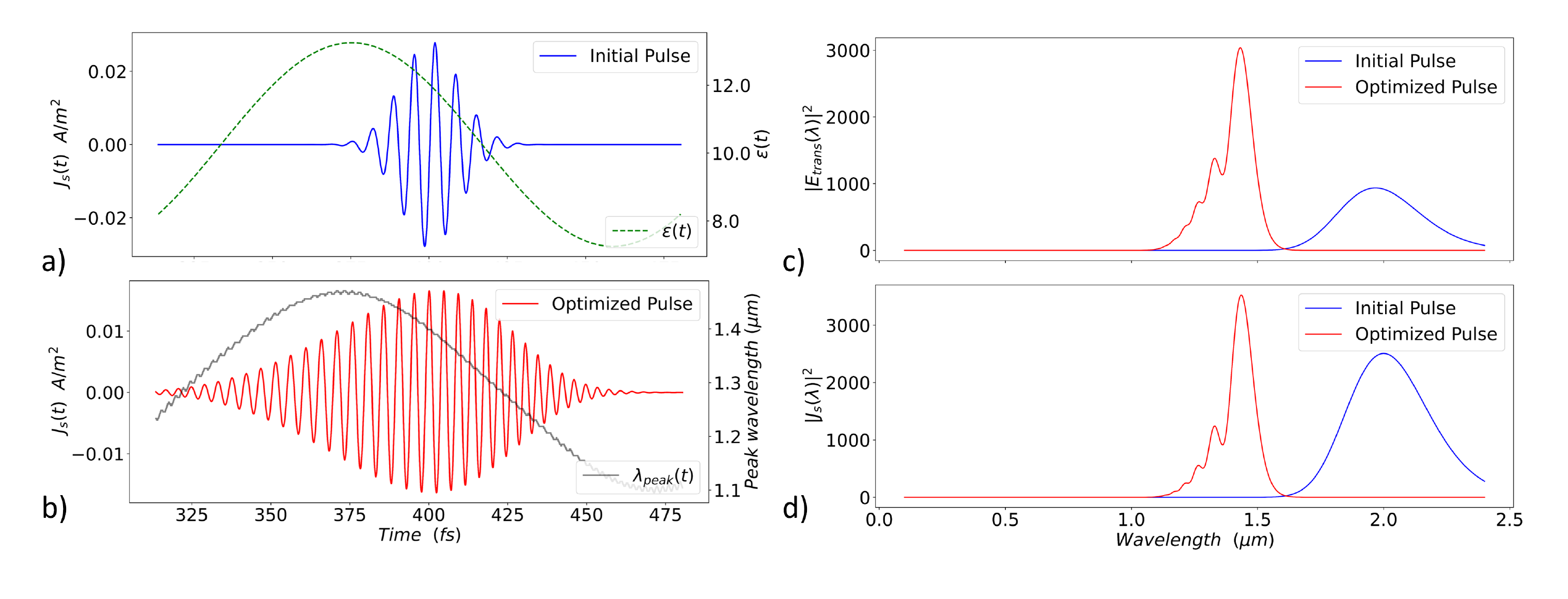}
\caption{a) Initial Gaussian input pulse, with duration 13.3 fs centered at 2 $\mu$m (blue line - left vertical axis) and the time-varying permittivity (green, dashed line - right vertical axis). b) Input pulse after optimization (red line - left vertical axis) and its instantaneous peak wavelength as a function of time (black line - right vertical axis). c) Transmission spectrum of the initial
 (blue line) and optimized (red line) pulses. d) Spectrum of the initial (blue line) and optimized (red line) pulses.}
\label{perm}
\end{figure}

As the first test case, we consider a 200 nm dispersionless thin film, with a slowly-varying sinusoidal modulation of the permittivity $\varepsilon(t)$, as plotted in Fig. \ref{perm}a (dashed green line – right axis) alongside the initial  pulse (solid blue line – left axis) that is taken to be a modulated Gaussian with a central wavelength of $\lambda_0 = 2$ $\mu m$ and duration $\tau = 13$ fs. The optimized pulse after 60 iterations is plotted in Fig. \ref{perm}b (red line - left axis). Note that in Figs. \ref{perm}a and b, only a portion of the pulses are shown for a small window of time in the larger simulation.
\newline

Also plotted in Fig. \ref{perm}b is the instantaneous peak wavelength of the optimized pulse as a function of time, calculated via windowed Fourier transforms (black line – right axis). It is clear that the optimized pulse is chirped in such a way that the instantaneous peak wavelength is following the time-varying permittivity of the thin-film, which would consequently have a time-varying transmission resonance. 
\newline

The transmitted energy spectra for the initial (blue line) and optimized (red line) pulses are plotted in Fig. \ref{perm}c, where we see that the transmitted pulse after optimization has a peak that is $3 \times$
higher than the initial pulse. The enhancement of the transmitted pulse energy after optimization (and thus the objective function $F$) is found by taking the ratio of the integral over the initial and optimized spectra, and is found to be larger than $2.5\times$. In Fig. \ref{perm}d we plot the spectra of the initial (blue line) and optimized (red line) pulses where we see that the optimized pulse is blue shifted, and the bandwidth is reduced. The spectrum is no longer a Gaussian, but includes additional frequency components. Note that the values on the left-hand axis for Figs. \ref{perm}c and \ref{perm}d are plotted such that they have the same units, that is, $J_s(\lambda)$ is scaled to have the same units as the electric field. This remains true in future sections, and makes physical interpretation easier. Indeed, comparing Fig. \ref{perm}c and Fig. \ref{perm}d, we find that the new pulse is almost completely transmitted, whereas the initial pulse is highly attenuated.

\subsection{Test case 2: Time-varying plasma frequency} \label{time_varying_drude1}
\begin{figure}
\centering
\includegraphics[width = \linewidth]{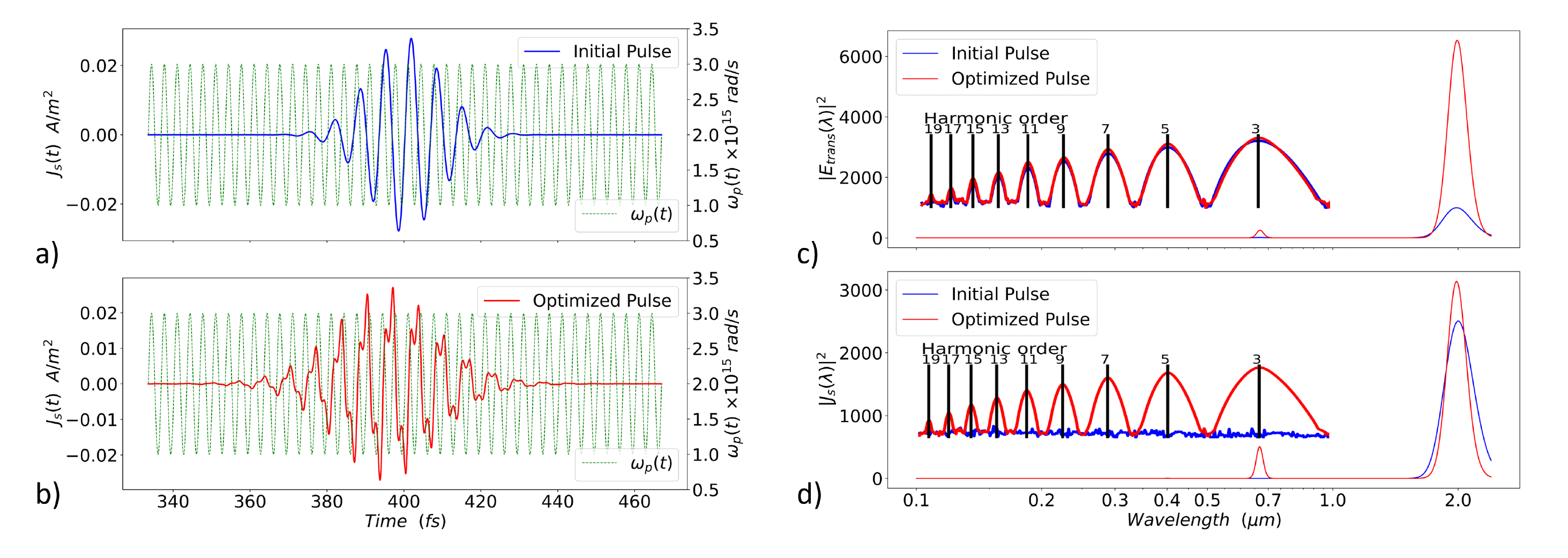}
\caption{a)Initial Gaussian input pulse with duration 13.3 fs centered at 2 $\mu$m (blue line - left vertical axis) and the time-varying plasma frequency (green, dashed line - right vertical axis). b) Input pulse after optimization (red line - left vertical axis) and the time- varying plasma frequency, shown again here for convenience (green, dashed
line - right vertical axis). c) Transmission spectrum of the initial (blue line) and optimized (red line) pulses. The inset shows a log plot at the lower wavelengths, demonstrating the generated odd harmonic orders, whose locations are indicated with black vertical lines. d) Spectrum of the initial (blue line) and the optimized (red line) pulses. The inset shows a log plot at the lower wavelengths, demonstrating the odd harmonic components in the optimized pulse only.}
\label{drude}
\end{figure}

For our second test case, we consider a 200 nm Drude metal thin film with a rapid sinusoidal time-dependent plasma frequency $\omega_p(t) \sim \sin(2 \omega_0 t)$, where $\omega_0$ is the central frequency of the initial pulse corresponding to $\lambda_0=2$ $\mu$m. The initial pulse is plotted in Fig. \ref{drude}a for a cropped time window, and is the same initial pulse shape as used in the previous section. The green dotted line shows the time-varying plasma frequency. After applying our optimization algorithm to the initial pulse in this time-varying medium, we obtain the optimized pulse shown in red in Fig. \ref{drude}b (in the same cropped time window as for panel a). In Fig. \ref{drude}c we plot the transmission spectrum of the initial (blue line) and optimized (red line) pulses; the inset shows the same in log scale.  The transmitted pulse energy from the optimized pulse has increased by a factor of 4.5 over that of the initial, unshaped pulse. Finally in Fig. \ref{drude}d we plot the pulse spectrum of the initial and optimized pulses, again with a log scaled-inset. 
\newline

The first effect of optimization on the input pulse is the increased pulse duration (see Fig. \ref{drude}b) and corresponding decrease in bandwidth near the central wavelength (see Fig. \ref{drude}d). The second effect is the appearance in Fig. \ref{drude}b of higher frequency components that modulate the pulse in time, that appear to sync with the modulated plasma frequency. From the optimized pulse spectrum in Fig. \ref{drude}d we see that these are odd harmonics. Indeed, it can be shown that odd harmonics are generated because the plasma frequency oscillates at $2\omega_0$ and are a demonstration of Floquet harmonics \cite{koutserimpas_parametric_2018}. Since these harmonics are transmitted out of the thin film (see the inset in Fig. \ref{drude}c), and because the thin film has a high transmittance at low wavelengths (a property of the Drude model), it is not a surprise that our optimization algorithm added these harmonics to our pulse (see Fig. \ref{drude}d). 
\newline

As mentioned in Section 3.3, the scale of the vertical axes of Figs. \ref{drude}c and d are in the same units, and we can see that the optimized transmitted pulse has experienced significant amplification. Indeed, due to the time-varying nature of these films, the energy of the input pulse need not be conserved, and gain is an expected consequence. It can be shown that parametric amplification occurs in periodic time-varying systems (time crystals) with a modulation frequency $\omega_{mod}=2\omega_0$, where $\omega_0$ is the frequency of light \cite{galiffi_photonics_2022,mendonca_temporal_2003,koutserimpas_parametric_2018}. The exponential gain is achieved due to the coherent sum of the forward scattered waves from the time- boundaries \cite{mendonca_temporal_2003,galiffi_photonics_2022}.
\newline 

Our pulse shaping method was able to find an optimized pulse that achieves broadband gain upon transmission through this time-varying Drude-metal. 
After integrating over both the initial and optimized transmission spectra, we find there is $\sim$ 2$\times$ more energy being transmitted in the optimized pulse. 

\section{Pulse shaping in frequency}

In this section we will introduce and demonstrate pulse shaping in frequency, wherein the phase of the frequency components is optimized, as would be in a phase-based 4f pulse shaper.  Unlike the pulse shaping in time method of the previous section, this method does not generate new frequency components so would be ideal for optimizing a pre-existing laboratory pulse of a set bandwidth. This method is naturally energy preserving, and the frequency spectrum remains the same for all possible pulse shapes (up to potentially small numerical errors arising from the discretization of the Fourier transform). If one were to consider amplitude-based pulse shaping (which would also be possible with our formalism, but that we do not consider here) then, of course, energy would not be conserved. 
\newline

Consider a current density $\bar{J}_s(t)$ that would represent the input to a 4f pulse shaper. The $j^{th}$ frequency component $\omega_j$ of this current source is found via a discrete Fourier transform of $\bar{J}_s(t)$,

\begin{equation} \label{source_freq}
    \bar{J}_s(\omega_j) = \sum_n \bar{J}_s(t_n)\exp(i\omega_jt_n).
\end{equation}

As in 4f pulse shaping, we allow each term in this sum to experience a distinct phase shift. We take these phase shifts $p_j = \phi(\omega_j)=\phi_j$ to be the tunable parameters in our optimization algorithm. We set the shaped pulse in time, and thus the source of our forward problems (Eqs. \ref{maxwell1} or \ref{maxwell2}) to be 

\begin{equation}
    J_s(t_n) = \frac{1}{2} \Bigg[\sum_j \bar{J}_s(\omega_j)\exp(-i\omega_j t_n + i\phi_j) + c.c.\Bigg]
\end{equation}

\noindent where c.c. means complex conjugate, which we have added to ensure $J_s(t_n)$ is real-valued. The gradient of the objective function $F$ with respect to the tuning parameters $p_j = \phi_j$ (Eq. \ref{grad3}) now becomes

\begin{equation}
    \frac{\partial F}{\partial \phi_j} = -\Delta t\sum_{n=1}^{N_T} E_x^{adj}(y_s,T_m-t_n) \text{Im}[\bar{J}_s(\omega_j)\exp(-i\omega_j t_n + i\phi_j)]
\end{equation}

\noindent where $N_T$ is the number of time iterations of our simulations. As in Section 3, we set the time step $\Delta t = t_{i+1}-t_i$  $\forall i$ as uniform. As before, only two simulations are required to calculate the gradient, which allows us to optimize $F$ with respect to the phase shifts $\phi_j$, and thereby tune the incident pulse.
\newline

In the following subsections, we will use our pulse shaping in frequency method to maximize the transmitted energy through time-varying dielectric and metallic thin films, similar to the previous section.

\subsection{Test case 1: Time-varying permittivity}
\begin{figure}
\centering
\includegraphics[width = \linewidth]{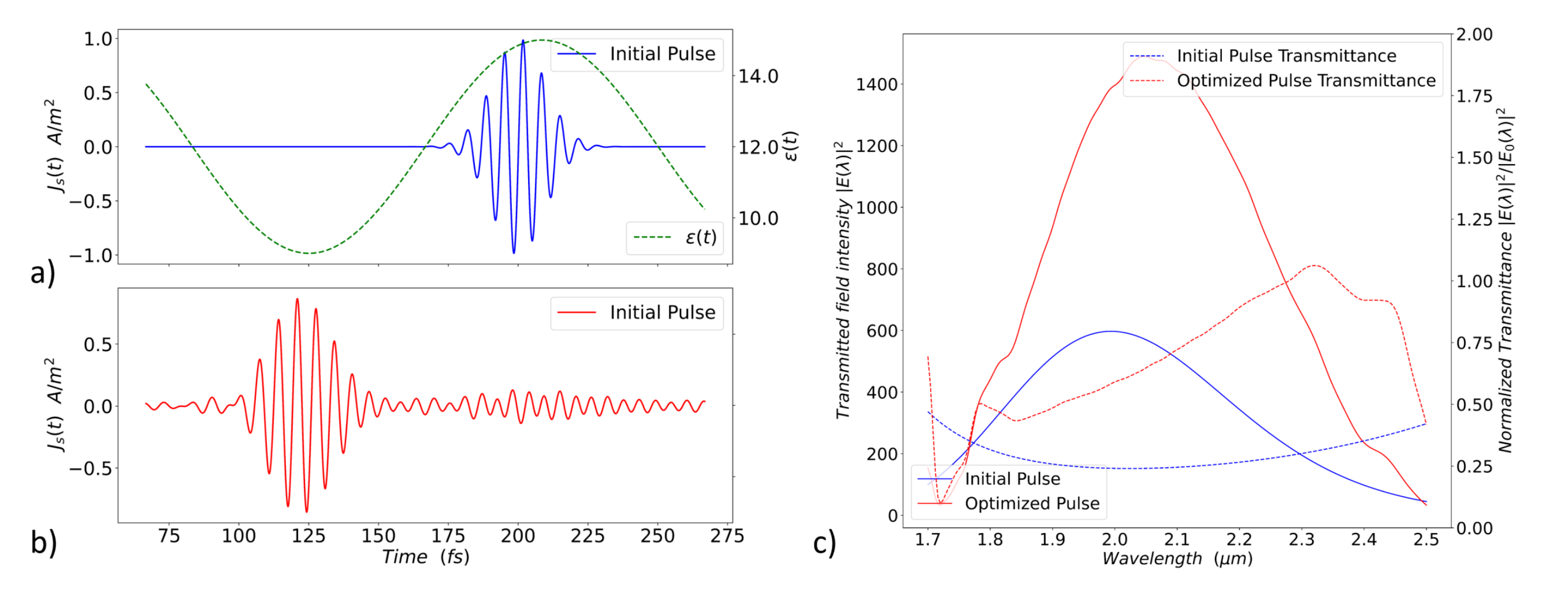}
\caption{a) Initial Gaussian input pulse with duration 13.3 fs centered at 2 $\mu$m (blue line - left vertical axis) and the time-varying permittivity oscillations (green, dashed line - right vertical axis). b) Optimized input pulse. c) Left vertical axis: transmission spectrum of the initial (blue line) and optimized (red line) pulses. Right vertical axis: normalized transmittance of the initial (blue, dashed line) and optimized (red, dashed line) pulses.}
\label{perm_freq}
\end{figure}

Here we test our pulse shaping in frequency method on a similar problem to that considered in Section 3.3, a 200 nm dispersionless thin film with a slowly-varying sinusoidal modulation $\varepsilon(t)$ as plotted in Fig. \ref{perm_freq}a (green, dashed line – right axis) alongside the initial pulse (solid blue line – left axis). The initial pulse is identical to that used in Section \ref{time_varying_perm1}. We use a numerical discrete Fourier transform to construct the pulse in the frequency domain via Eq. \ref{source_freq} with 300 discrete frequencies sampled evenly between $0.5\omega_0$ and $2\omega_0$, where $\omega_0$ is the pulse center frequency.
\newline

The optimized pulse after 60 iterations is plotted in Fig. \ref{perm_freq}b. The main effect of the optimization is to create a delay such that the majority of the pulse has been shifted to where the permittivity is lowest (around 125 fs) as the instantaneous transmittance would be highest at this permittivity minima. In fact, near the minima of the permittivity where $\varepsilon \sim 9$, the thin film has a resonance around $\lambda \sim 2400$ nm. As the permittivity increases, the resonance wavelength will further redshift from the bandwidth of the pulse (which, recall, is centred at 2 $\mu$m). Indeed, the best course of action for the optimizer was to move the pulse to a lower permittivity, where the resonance overlaps with the pulse bandwidth.
\newline

In Fig. \ref{perm_freq}c we plot the transmitted intensity (left vertical axis – solid lines) and the normalized transmittance (ratio of transmitted to input intensity, right vertical axis – dashed lines) for the initial (blue line) and optimized (red line) pulses. We have achieved more than 2.4 $\times$ enhancement of the transmitted intensity at the peak wavelength, and 2.2$\times$ enhancement in transmitted energy (that is, integrated across the spectrum).  Here we see further evidence for our physical interpretation of the action of the optimization on the pulse. We see that enhancement in transmittance (comparing the dashed red and blue lines) is highly biased towards the higher (resonant) wavelengths. A notable feature is the presence of gain at around 2320 nm. This is most likely due to frequency translation owing to the time-varying permittivity \cite{xiao_spectral_2011}. This frequency translation causes the pulse frequency components to shift while traversing a time-varying material. In spectral regions of high transmittance, frequency translation can result in a transmittance > 1.
\newline

From this simple model we can see that by selectively delaying the frequency components of the pulse, we can achieve a broadband transmittance enhancement across the spectrum of the pulse without changing the amplitude of the pulse spectral components. 

\subsection{Test case 2: Time-varying plasma frequency}

\begin{figure}
\centering
\includegraphics[width = \linewidth]{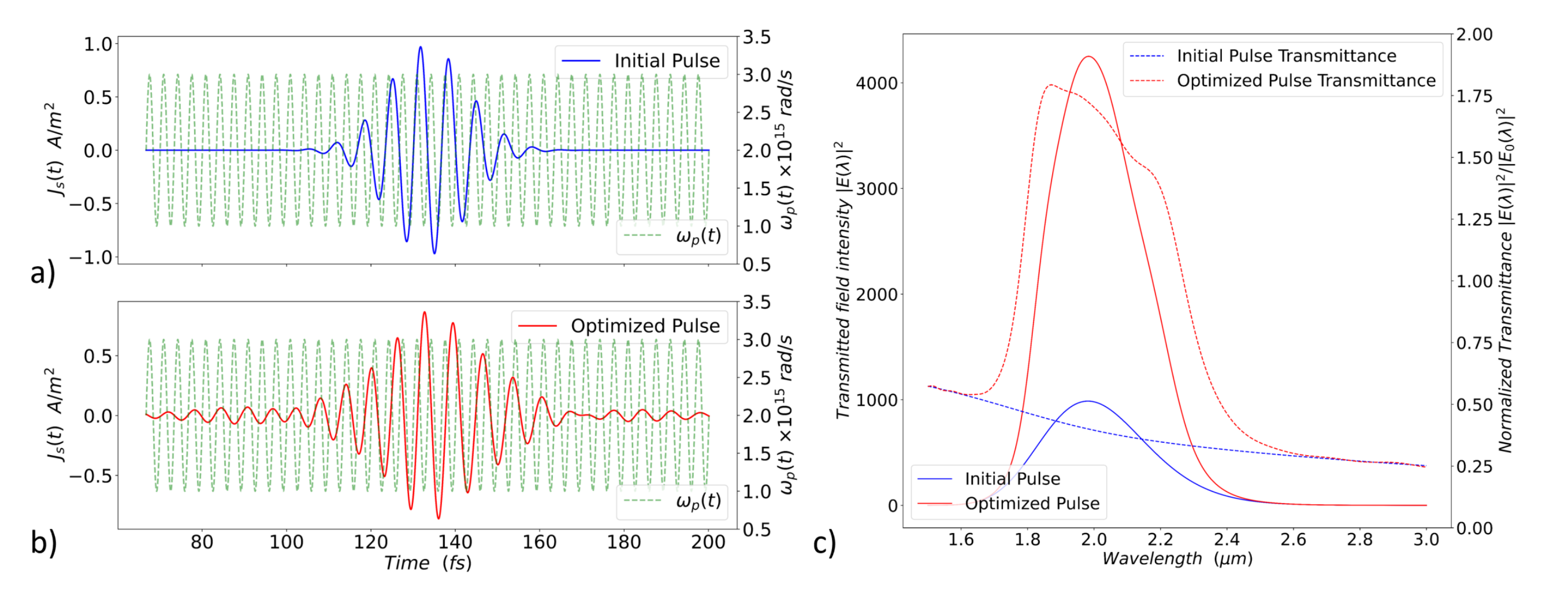}
\caption{a) Initial Gaussian input pulse with duration 13.3 fs centered at 2 $\mu$m (blue line - left vertical axis) and the time-varying plasma frequency (green, dashed line - right vertical axis). b) Input pulse after optimization and time-varying plasma frequency shown again for convenience (green dashed line - right vertical axis). c) Left vertical axis: transmission spectrum of the initial (blue line) and optimized (red line) pulses . Right vertical axis: normalized transmittance of the initial (blue, dashed line) and optimized (red, dashed line) pulses.}
\label{drude_freq}
\end{figure}

In this example, we use a similar setup as in Section \ref{time_varying_drude1}. We use the same initial pulse (plotted in Fig. \ref{drude_freq}a, blue line)  incident on a 200 nm Drude-metal film with the same time-varying plasma frequency as before (plotted in Figs. \ref{drude_freq}a and b, green dashed line)
\newline

The optimized pulse after 30 iterations is plotted in Fig. \ref{drude_freq}b. We see the optimized pulse (red line) is largely unchanged, but slightly delayed relative to the initial pulse (Fig. \ref{drude_freq}a, blue line). This delay works to align the pulse sub-cycle peaks with the plasma frequency troughs (green-dashed line in Figs. \ref{drude_freq}a and b) near the middle of the pulse. The optimized pulse is also stretched in time.
\newline

In Fig. \ref{drude_freq}c we plot the transmission spectrum (left vertical axis – solid lines) and the normalized transmittance (right vertical axis – dashed lines) of the initial (blue line) and optimized (red line) pulses. As in Section \ref{time_varying_drude1}, we see that our new pulse has again achieved gain due to parametric amplification. We achieve a 3.5$\times$ increase in the transmitted energy (that is, integrated across the transmission spectrum). Once again we show how pulse shaping can be used to achieve broadband gain via transmission through time-varying media, only this time through the controlled delay of the pulse frequency components. 

\section{Pulse shaping for a strongly pumped, ENZ material}

In the previous section, we explored our pulse shaping inverse design processes by maximizing the energy transmitted through time-varying materials based on toy-models. In this section, we will demonstrate both our time and frequency pulse shaping methods for a realistic time-varying medium by maximizing the transmitted energy of a probe pulse through a pumped ITO thin film.
\newline

%ITO’s 
The permittivity of ITO is highly dependent on the temperature of the conduction band electrons, and as such, it is a time-varying material under ultrafast pulse irradiation, especially near its ENZ band \cite{alam_large_2016,alam_large_2018,zhou_broadband_2020,reshef_beyond_2017}. Because it exhibits strong, and fast permittivity perturbations, ITO is a material of high interest in the field of active nanophotonics. Its nonlinear optical properties are well studied and can be modelled using a self-consistent multiphysics model that couples electrodynamics and thermodynamics introduced in Ref. \cite{baxter_dynamic_nodate}.

\subsection{Pump simulation}

\begin{figure}
\centering
\includegraphics[width = 0.7\linewidth]{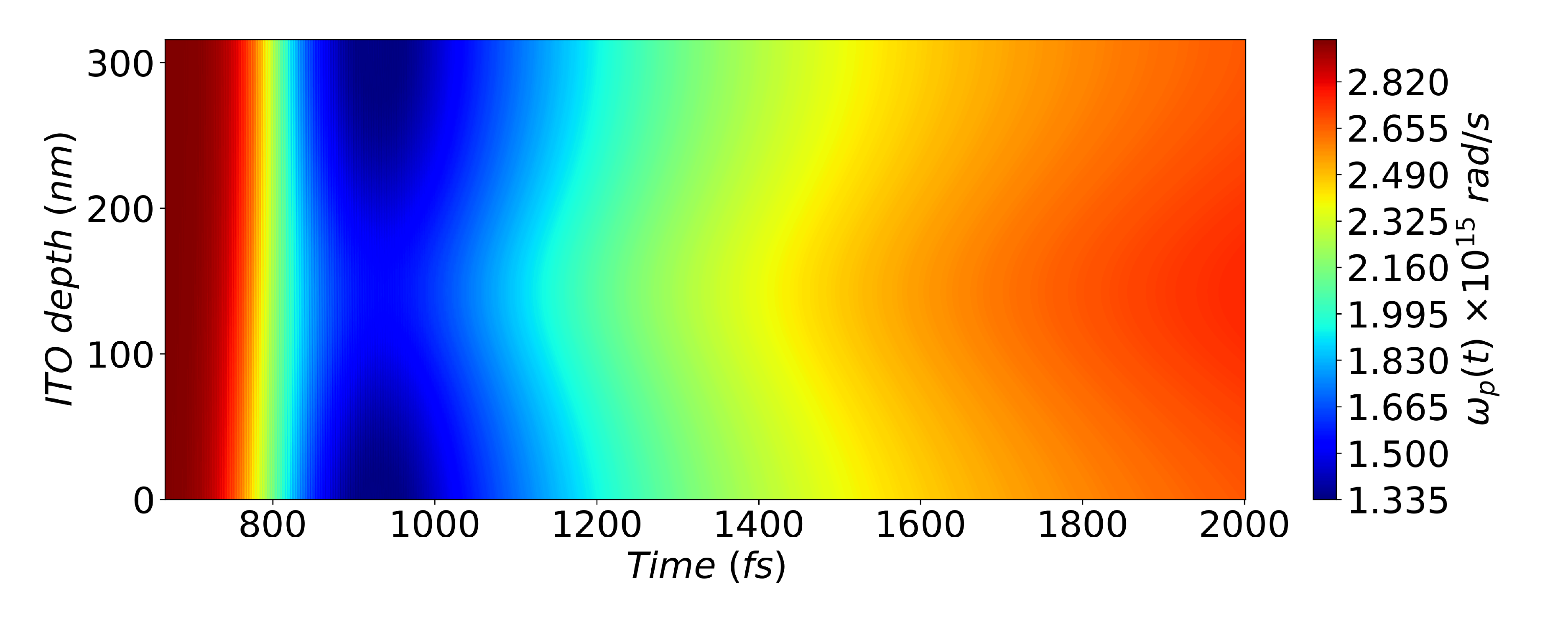}
\caption{Colour contour plot of the space (vertical axis) and time (horizontal axis)-varying plasma frequency $\omega_p (y,t)$ which is extracted from the multiphysics modelling of an ITO thin film under high intensity pulse irradiation \cite{baxter_dynamic_nodate}.}
\label{plasma_freq}
\end{figure}

Our goal will be to optimize a probe pulse incident on an ITO thin film after the film has been irradiated by an ultrafast intense light pulse, which we call the pump pulse. This pump pulse creates a time-varying medium through a temperature dependent plasma frequency, that we simulate by implementing the model of Ref. \cite{baxter_dynamic_nodate} into a 1D-FDTD solver (with the code provided as supplemental material). We simulate a modulated Gaussian pump pulse with peak intensity $I_{peak}=13.3$ TW/cm\textsuperscript{2}, pulse duration $\tau=100$ fs, and center wavelength $\lambda_0=1.23$ $\mu$m incident on a 320 nm thin film of ITO. This center wavelength corresponds to the ENZ wavelength of the ITO film. We store the time and spatially dependent plasma frequency $\omega_p (\textbf{r},t)$ at each time-step and position in a text file for future use; this could be cumbersome for 2 and 3D geometries if a large number of points are required. We plot $\omega_p (\textbf{r},t)$ in Fig. 7 where the vertical axis represents the depth in the ITO film (where $\textbf{r}=y\hat{\mathbf{y}}$), and the horizontal axis represents time (which is cropped to highlight the important time window). The unpumped ITO film has a plasma frequency everywhere of
$\omega_p=2.97\times10^{15}$ rad/s. Upon irradiation with the pump pulse, the plasma frequency in the ITO decreases as the excited conduction electrons occupy higher energy levels \cite{alam_large_2016,baxter_dynamic_nodate,guo_ultrafast_2016}. In Fig. \ref{plasma_freq}, the plasma frequency drops to half of its initial value through most of the ITO layer in the period of $\sim 200$ fs and it's return to equilibrium will take several picoseconds (not shown in figure).
\newline

With the spatiotemporal plasma frequency of the pumped ITO film stored, and the time-varying medium thus defined, we can now apply our time and frequency pulse shaping techniques to maximize the transmitted energy of a probe pulse incident on the pumped ITO film. We turn to this in the next two subsections.

\subsection{Probe optimization: Pulse shaping in time}

\begin{figure}
\centering
\includegraphics[width = \linewidth]{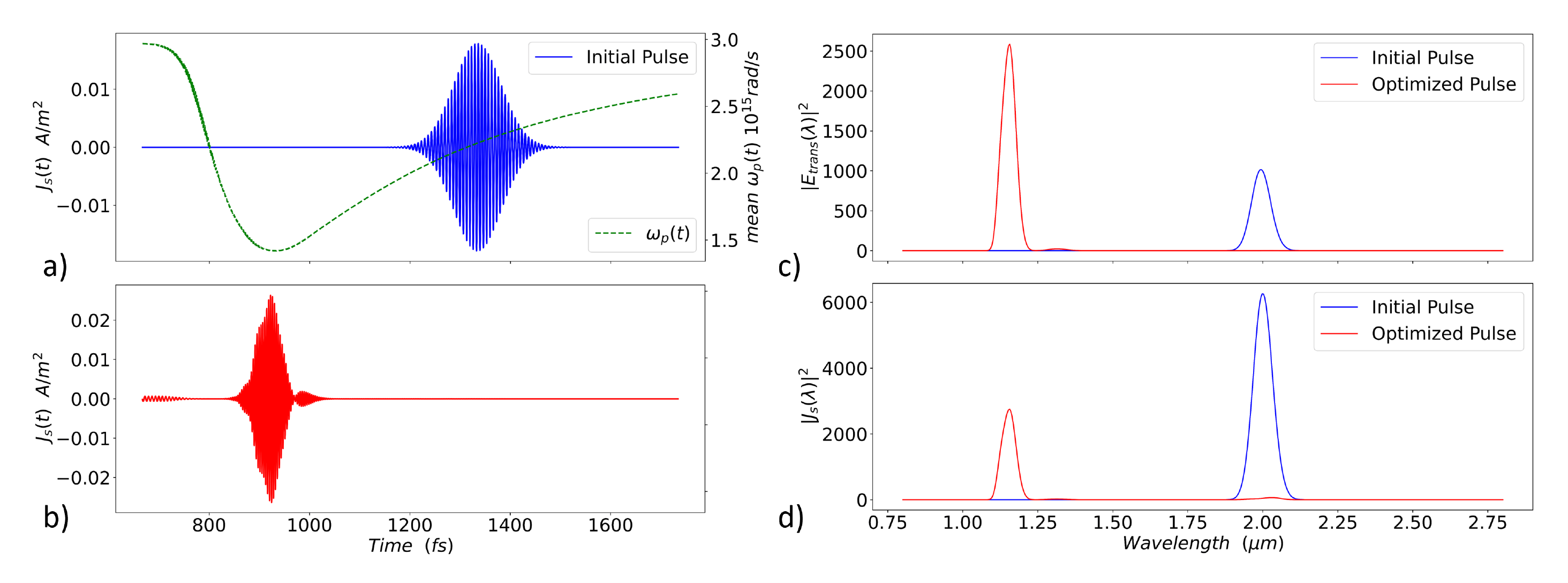}
\caption{a) Initial Gaussian probe pulse with duration 67 fs centered at 2 $\mu$m (blue line) and, b) optimized probe pulse (red line). In a) we overlay the spatially-averaged plasma frequency (dashed green line). c) Transmission spectrum of the initial (blue line) and optimized (red line) pulses. d) Spectrum of the initial (blue line) and optimized (red line) pulses.}
\label{ITO}
\end{figure}

In this subsection, we will use our pulse shaping in time method to optimize the transmitted energy of the probe pulse incident on the pumped ITO film. The initial probe pulse is a Gaussian centered at $\lambda_0=2$ $\mu$m with pulse duration $\tau=67$ fs. It is plotted in Fig. \ref{ITO}a (blue line) alongside the space-averaged plasma frequency from Fig. \ref{plasma_freq} for reference (dashed, green line). The optimized pulse after 300 iterations is plotted in Fig. \ref{ITO}b.
\newline

Additional filtering was required to keep the pulse spectrum in the near-infrared as the optimizer prefers to introduce frequency components as high as possible to capitalize on the high-frequency transparency of metals. Prior to the forward simulation, the (pre-filtered) pulse $p(t)$ is band-pass filtered via fast-Fourier transforms. As in Section 3, the pulse is normalized to ensure no energy is added or removed from the filtered pulse $J_s(t)$. This filtering procedure is differentiable via automatic differentiation \cite{noauthor_autograd_2022}. The derivative $\frac{\partial J_s(t)}{\partial p_i}$ in Eq. \ref{grad3} now accounts for the band-pass filtering and normalization, where $p_i$ is the $i^{th}$ timestep of the pre-filtered pulse $p(t)$ that we are optimizing. Although not necessary, automatic differentiation can also be applied to the pulse normalization of Section 3.
\newline

The major effect of the optimization is to shift the pulse in time, such that most of the energy is near the minimum of the plasma frequency. As the plasma frequency decreases, the imaginary component of the permittivity $Im(\varepsilon(\omega))$ also decreases for $\omega > \omega_p$, thus reducing the loss, and the real component $Re(\varepsilon(\omega))$ increases. The film exhibits dielectric behaviour, and it is understandable why the optimizer chose to delay the pulse as it did, and reduce the center wavelength. Furthermore, as plotted in Fig. \ref{ITO}c, the optimized pulse is centered at $\lambda_{0,opt} = 1177$ nm corresponding to a transmittance maximum for a thin film with a static plasma frequency of $\omega_p=1.5\times10^{15}$ rad/s. The transmitted spectra are plotted in Fig. \ref{ITO}d where the energy of the optimized pulse is 92\% transmitted, $5.5\times$ higher than the initial pulse. 

\subsection{Probe optimization: Pulse shaping in frequency}

\begin{figure}
\centering
\includegraphics[width = \linewidth]{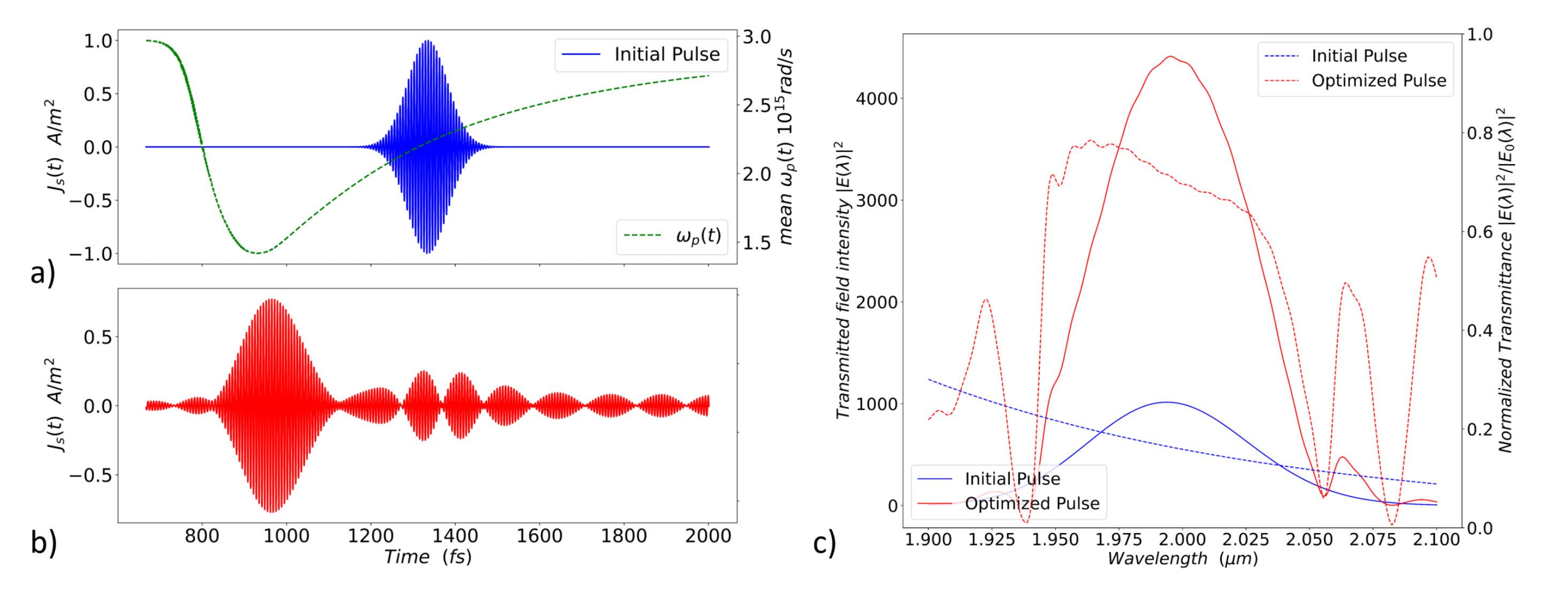}
\caption{a) %Original 
Initial Gaussian probe pulse with duration 67 fs centered at 2 $\mu$m (blue line) and, b) optimized probe pulse %haped pulse after optimization 
(red line). In a) we overlay the spatially-averaged plasma frequency (dashed green line) for comparison. c) Left vertical axis: transmission spectrum of the initial %original 
pulse (blue line) and the optimized pulse (red line). Right vertical axis: Normalized transmittance of the initial (blue, dashed line) and optimized (red, dashed line) pulses.}
\label{ITO_freq}
\end{figure}

Like in the previous section, here we are optimizing the transmitted energy of the probe pulse traversing the pumped ITO film simulated in Section 5.1, only this time using our pulse shaping in frequency method. The initial probe pulse is again a Gaussian centered at $\lambda_0=2$ $\mu$m with  duration $\tau=67$ fs, as plotted in Fig. \ref{ITO_freq}a (blue line) along with the space-averaged plasma frequency from the pump simulation (dashed, green line) calculated from Fig. \ref{plasma_freq}. 
\newline

After 30 iterations of pulse shaping, we obtain the optimized pulse plotted in Fig. \ref{ITO_freq}b. Once again, we see the pulse being time-shifted such that it overlaps in time with the plasma frequency minimum. In Fig. \ref{ITO_freq}c we plot the transmitted field intensity (left vertical axis – solid lines) and the normalized transmittance (right vertical axis – dashed lines) of the initial (blue line) and optimized (red line) pulses. The transmitted energy is increased by a factor of 4$\times$ after optimization. 
\newline

\begin{figure}
\centering
\includegraphics[width = 0.5\linewidth]{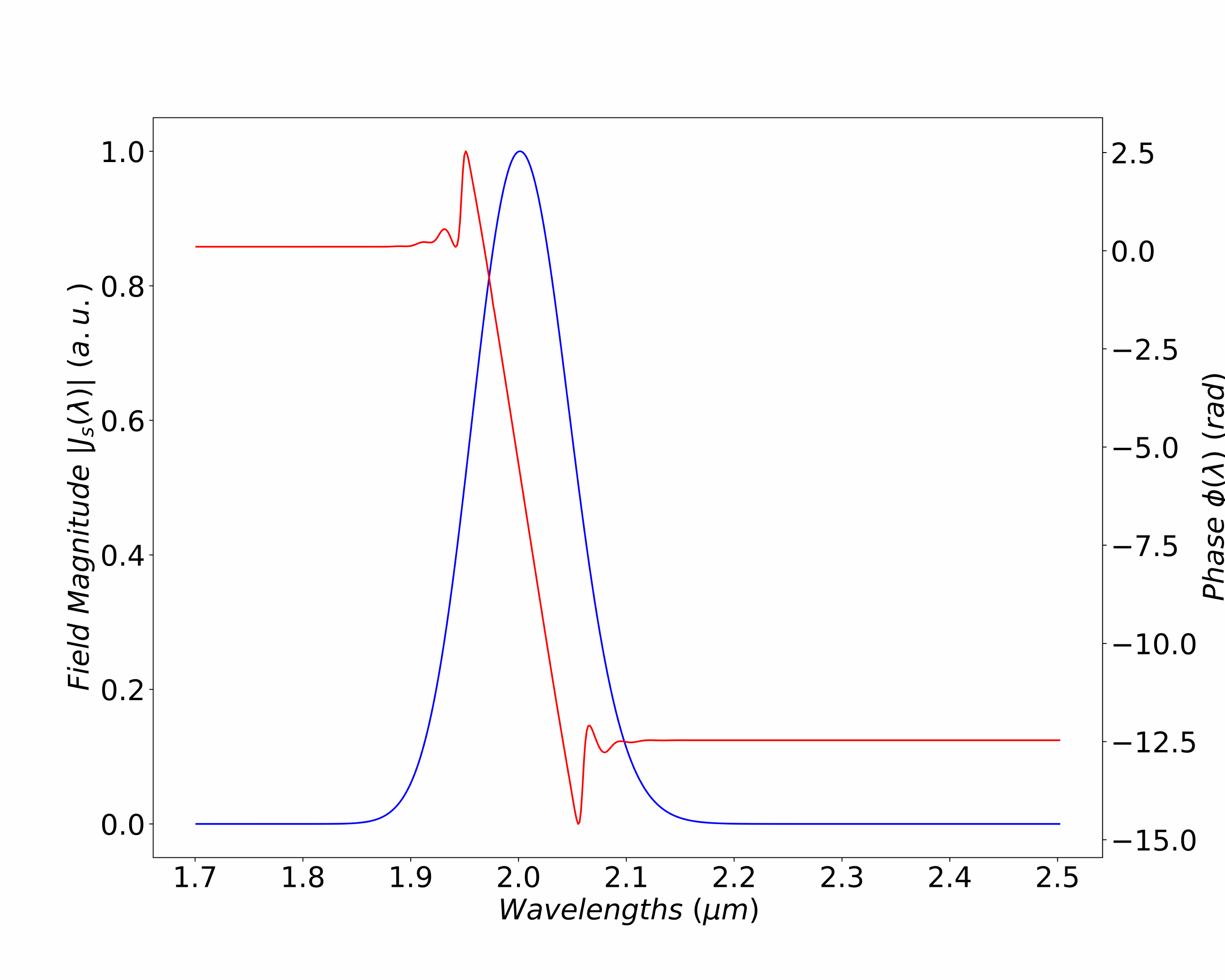}
\caption{Optimized probe pulse spectrum. Left vertical axis: pulse amplitude (blue line). Right vertical axis: pulse phase (red line)}
\label{ITO_freq_spectrum}
\end{figure}

In Fig. \ref{ITO_freq_spectrum} we plot the optimized pulse spectrum (blue line) and optimized phase (red line). The main effect of the pulse shaping is to linearly decrease the phase %linear phase decrease  with  increasing wavelength.
as a function of wavelength.  The time delay of a given frequency component in a phase-shaped pulse is given by $\tau(\omega)=-\partial\phi(\omega)/\partial\omega$ \cite{weiner_ultrafast_2011} which is constant when $\phi(\omega)$ is linear. A quick calculation reveals that most frequency components in the pulse bandwidth are delayed by $\tau \approx -370$ fs, which is apparent in Fig. \ref{ITO_freq}b.

\section{Conclusion}
We have introduced a method for the inverse design of optical pulse shapes for time-varying nanophotonic systems, opening a new paradigm for control over light-matter interaction. We derive the sensitivity of an objective function to the pulse shape in two ways. First, we introduce pulse shaping in time, where the gradient of the objective function with respect to the pulse amplitude at a given time can be extracted using two FDTD simulations, allowing for pulses of arbitrary frequency components (but same pulse energy) to be designed. Second, we develop pulse shaping in frequency, where the phase of discrete frequencies in the pulse are tuned, replicating a 4f pulse-shaping setup. We demonstrate these methods in time varying materials, including the optimization of a probe pulse in intensely irradiated ITO thin films.  This is an unconventional approach to computational-design in nanophotonics, but one that is likely to be important in the near-future given the current interest in active nanophotonics.

\newpage

\renewcommand{\theequation}{S.\arabic{equation}}

\section{Supporting Information}
\subsection{Adjoint method in the time domain}\label{td-adjoint}

In this section, we provide a derivation of the adjoint method in the time domain for completeness, following Ref. \cite{bakr_adjoint_2017}. We begin with Maxwell's equations, which can be written as, 

\begin{equation} \label{maxwell_general}
    A\dot{\textbf{x}} + B\textbf{x} = \textbf{s}
\end{equation}

\noindent where $A$ and $B$ are the system matrices, $\textbf{x}$ is the electromagnetic field vector, and $\textbf{s}$ is the source vector. In order to obtain the gradient, we differentiate Eq. \ref{maxwell_general} with respect to some tunable parameter $p_i$,

\begin{equation} 
    \frac{\partial A}{\partial p_i}\dot{\textbf{x}} + A\frac{\partial \dot{\textbf{x}}}{\partial p_i} + \frac{\partial B}{\partial p_i}\textbf{x} + B\frac{\partial \textbf{x}}{\partial p_i} = \frac{\partial \textbf{s}}{\partial p_i}.
\end{equation}

\noindent In order to scalarize this equation, we left-multiply by an adjoint electromagnetic field $\textbf{x}_{adj}$ vector (to be specified later),

\begin{equation} \label{intermediate}
    \textbf{x}_{adj}^T A\frac{\partial \dot{\textbf{x}}}{\partial p_i} + \textbf{x}_{adj}^T B\frac{\partial \textbf{x}}{\partial p_i} = \textbf{x}_{adj}^T \textbf{R}_i,
\end{equation}

\noindent where we have introduced the residue vector

\begin{equation}
    \textbf{R}_i = \frac{\partial \textbf{s}}{\partial p_i} - \frac{\partial A}{\partial p_i}\dot{\textbf{x}} - \frac{\partial B}{\partial p_i}\textbf{x}.
\end{equation}

\noindent Finally, we time-integrate Eq. \ref{intermediate} over the simulation ($t = 0$ to $t=T_m$), 

\begin{equation}\label{integrand}
    \int_0^{T_m} (-\dot{\textbf{x}}_{adj}^T A + \textbf{x}_{adj}^T B) \frac{\partial\textbf{x}}{\partial p_i} dt = \int_0 ^{T_m} \textbf{x}_{adj}^T \textbf{R}_i dt.   
\end{equation}

\noindent making the assumption that $\textbf{x}$ and $\textbf{x}_{adj}$ are negligible outside this time window. The bracketed expression within the left-hand integrand resembles the left hand side Maxwell's equation \ref{maxwell_general}, but here for the adjoint field. Taking the transpose of the bracketed term, and setting it equal to $\partial\psi / \partial \textbf{x}$, we obtain 

\begin{equation} \label{roughadjoint}
    -A^T \dot{\textbf{x}}_{adj} +  B^T \textbf{x}_{adj} = \frac{\partial\psi}{ \partial \textbf{x}},
\end{equation}

\noindent then Eq. \ref{integrand} becomes

\begin{equation}\label{integrand2}
    \int_0^{T_m}  \Big(\frac{\partial\psi}{ \partial \textbf{x}}
    \Big)^T \frac{\partial\textbf{x}}{\partial p_i} dt = \int_0 ^{T_m} \textbf{x}_{adj}^T \textbf{R}_i dt.   
\end{equation}

\noindent From Eq. \ref{integrand2}, we see that the integrand on the left hand side is simply the chain-rule expansion of $\partial \psi /\partial p_i$ and thus we have our gradient of the objective function defined in Eq. \ref{generalgrad},

\begin{equation}\label{formal_gradient}
    \frac{\partial F}{\partial p_i} = \frac{\partial}{\partial p_i}   \int_0^{T_m}  \psi(\textbf{x},t) dt = \int_0 ^{T_m} \textbf{x}_{adj}^T \textbf{R}_i dt.
\end{equation}

We can calculate $\textbf{R}_i$ using forward simulation data obtained by simulating \ref{maxwell_general}. The adjoint field is calculated using the time reversal of Eq. \ref{roughadjoint}

\begin{equation} \label{General_adjoint}
    A^T \dot{\textbf{x}}_{adj} +  B^T \textbf{x}_{adj} = \frac{\partial\psi}{ \partial \textbf{x}}(T_m-t),
\end{equation}

\noindent which ensures the sign of each term has the same sign as Eq. \ref{maxwell_general}. Note that if $A$ and $B$ are time-varying they must also be time-reversed.
\newline

Eq. \ref{General_adjoint} is the general formula for the adjoint simulations. For most static, optical modelling, $A$ is diagonal and $B$ is symmetric, and thus the adjoint simulator is identical to the forward simulator with a different source. However, for more complicated problems (as we shall see in the following subsections), the matrices are not symmetric and the simulation software must be adapted.

\subsubsection{Time-varying permittivity}\label{2.1}
Let us consider a non-magnetic medium with a permittivity that can vary in both space and time but is otherwise dispersionless. In this case, Maxwell’s equations (Eq. \ref{maxwell_general}) can be written as 

\begin{gather} \label{maxwell1}
 \begin{bmatrix} \varepsilon_0 \varepsilon(t) & 0 \\ 0 & -\mu_0 \end{bmatrix}
\frac{\partial}{\partial t} \begin{bmatrix} \mathbf{E} \\ \mathbf{H} \end{bmatrix}
+ \begin{bmatrix} \varepsilon_0 \dot{\varepsilon}(t) & -\nabla\times \\ -\nabla\times & 0\end{bmatrix}
\begin{bmatrix} \mathbf{E} \\ \mathbf{H} \end{bmatrix} = \begin{bmatrix} -\mathbf{J}_s(t) \\ \mathbf{M}_s(t) \end{bmatrix},
\end{gather}

\noindent Implicit in this algebraic notation is the assumption that Eq. \ref{maxwell1} will be solved on a discretized spatial grid with, say, $N$ cells. Then the electric ($\textbf{E}$) and magnetic ($\textbf{H}$) fields are themselves column vectors of length $3N$,  where the factor of $3$ comes from the three components of  $\textbf{E}$ and $\textbf{H}$; the matrix elements (ie. $\varepsilon_0\varepsilon(t)$, $\varepsilon_0 \dot{\varepsilon}(t)$, $-\mu_0$, $0$, and $-\nabla \times$) are themselves $3N\times3N$ sub-matrices. Since we are considering non-magnetic materials, the $-\mu_0$  sub-matrix is diagonal, with each diagonal element having the value of the (negative of the) permeability of free space. The permittivity sub-matrix $\varepsilon_0 \varepsilon(t)$ need not be diagonal, but in our examples below we assume isotropic materials, so it is taken as diagonal. The sub-matrix $\nabla\times$ is of the form

\begin{gather}
   \nabla\times\textbf{E} =  \begin{bmatrix} 0 & -\partial/\partial z & \partial/\partial y \\ 
   \partial/\partial z & 0 & -\partial/\partial x \\ 
   -\partial/\partial y & \partial/\partial x & 0 \end{bmatrix}   \begin{bmatrix}
       E_x \\ E_y \\ E_z 
   \end{bmatrix},
\end{gather}

\noindent where each partial derivative is a $N\times N$ finite difference matrix. The $\varepsilon_0 \dot{\varepsilon}(t)$ sub-matrix arises naturally in Maxwell’s equations for a time-dependent permittivity.
\newline

The quantities that define the input pulse are $\textbf{J}_s$ and $\textbf{M}_s$, the time-dependent external current density and magnetization sources, respectively, that are both represented as column vectors of length $3N$. It is the shape of $\textbf{J}_s$ and/or $\textbf{M}_s$ that we seek to tune in order to maximize our objective. Note that in the following, we will assume the external source is non-magnetic, so we set $\textbf{M}_s=0$.
\newline

The corresponding adjoint matrix equation (Eq. \ref{General_adjoint}) is given by

\begin{gather} \label{adjoint1}
 \begin{bmatrix} \varepsilon_0 \varepsilon(T_m-t) & 0 \\ 0 & -\mu_0 \end{bmatrix}
\frac{\partial}{\partial t} \begin{bmatrix} \mathbf{E}^{adj} \\ \mathbf{H}^{adj} \end{bmatrix}
+\begin{bmatrix} -\varepsilon_0 \dot{\varepsilon}(T_m-t) & -\nabla\times \\ -\nabla\times &  0 \end{bmatrix}
\begin{bmatrix} \mathbf{E}^{adj} \\ \mathbf{H}^{adj} \end{bmatrix} = \begin{bmatrix} -\mathbf{J}^{adj}_s(t) \\ 0 \end{bmatrix},
\end{gather}

\noindent where the external source is taken to be

\begin{equation} \label{adjointsource}
    \textbf{J}_s^{adj}=-\frac{\partial\psi}{\partial\textbf{E}} (T_m-t). 
\end{equation}

\noindent Here we have assumed that $\psi$ depends only on $\textbf{E}$. Since we choose the functional form of $\psi(\textbf{E})$, we can derive an analytic expression for $\partial\psi/\partial\textbf{E}$ as a function of $\textbf{E}$, where $\textbf{E}$ is to be calculated from Eq. \ref{maxwell1} (the forward simulation). For general objective functions, $\psi$ could also be a function of $\textbf{H}$ and thus the adjoint problem may also contain a magnetic source $\textbf{M}_s^{adj}=\frac{\partial\psi}{\partial\textbf{E}} (T_m-t)$, though we do not consider this here. Recent implementations of adjoint sensitivity analysis use automatic differentiation to calculate $\partial\psi/\partial\textbf{E}$ and $\partial\psi/\partial\textbf{H}$  for user-defined objective functions \cite{hammond_high-performance_2022,hughes_forward-mode_2019}.    
\newline

For the case where $\varepsilon$ is not time-varying, $\dot{\varepsilon}=0$, and the adjoint problem defined by Eq. \ref{adjoint1} is identical to the forward problem of Eq. \ref{maxwell1}, the only difference being the external current density source in the adjoint simulation is given by \ref{adjointsource}. However, in time-varying materials, the adjoint problem is not the same as the forward one, as $\varepsilon(t)$ is time-reversed and $\dot{\varepsilon}$ is time-reversed and changes sign. This requires a small but straightforward change to the simulation software.
\newline

Once the forward and adjoint problems are computed, one can use the forward and adjoint fields to compute the gradient, Eq. \ref{formal_gradient} via

\begin{gather} \label{grad}
 \frac{\partial F}{\partial p_i}= \int_0^{T_m}\begin{bmatrix} \mathbf{E}^{adj}(T_m-t) \\ \mathbf{H}^{adj}(T_m-t) \end{bmatrix} \cdot \textbf{R}_i(t) dt,
\end{gather}

\noindent where the integrand is the scalar product of the time-reversed adjoint fields and the residue vector

\begin{gather} \label{residuefull}
\textbf{R}_i(t) = -\Bigg(\frac{\partial}{\partial p_i} \begin{bmatrix} \varepsilon_0 \varepsilon(t) & 0 \\ 0 & -\mu_0 \end{bmatrix}\Bigg)\frac{\partial}{\partial t} \begin{bmatrix} \mathbf{E}(t) \\ \mathbf{H}(t) \end{bmatrix}
+\Bigg(\frac{\partial}{\partial p_i}\begin{bmatrix} -\varepsilon_0 \dot{\varepsilon}(t) & \nabla\times \\ \nabla\times & 0\end{bmatrix}\Bigg)
\begin{bmatrix} \mathbf{E}(t) \\ \mathbf{H}(t) \end{bmatrix} + \frac{\partial}{\partial p_i}\begin{bmatrix} -\mathbf{J}_s(t) \\ 0 \end{bmatrix}.
\end{gather}

Note that this formalism allows for simultaneous optimization of both input pulse shape and topology (the latter being somewhat computationally expensive in the time-domain). For topology optimization, where we tune $\varepsilon$ over many spatial grid cells, only the first term is required for a static $\varepsilon$, but both the first and second terms are required for a time-dependent $\varepsilon$. However, for the optimization of pulse shape for a given topology, we need only consider the third term. As this is our interest here, we thus set 

\begin{gather} \label{residue}
\textbf{R}_i(t) = \frac{\partial}{\partial p_i}\begin{bmatrix} -\mathbf{J}_s(t) \\ 0 \end{bmatrix},
\end{gather}

\noindent so that no forward fields are required to obtain the residue, unlike for topology optimization. However, the adjoint fields (in particular, the electric adjoint field) are required at the spatial locations from which the source $\textbf{J}_s(t)$ is injected.
\newline

In time-domain topology optimization, the time-domain fields must be kept at all spatial locations over which topology optimization is to occur, typically resulting in large memory and I/O requirements that can present a limitation, especially for simulations involving long time-domain signals. The time-domain fields are required for the calculation of the first two terms of the residue in Eq. \ref{residuefull}, and are in general over a 3D volume \cite{hassan_topology_2022}. 
\newline

For pulse shaping, in contrast, fields are not required for the residue $\textbf{R}_i (t)$ calculation. Rather, $\textbf{R}_i (t)$ only depends on the external source, which can be further parameterized and is known analytically. Thus, the only field data that absolutely needs to be kept from the forward simulation are the fields at the locations at which the objective function is defined. If this happens to be only a single or few points, then the memory and I/O requirements are, in fact, modest. If the objective function is defined over a significant number of points, then memory and I/O requirements may again start to present a limitation, though not as severely as for 3D time-domain topology optimization; typically objective functions are defined on at most a 2D sub-space, such as a transmission plane.

\subsubsection{Time-varying dispersion}\label{2.2}

Materials with time-varying optical properties can also be described by dispersive models, such as the Drude and Lorentz models, and many others \cite{taflove_computational_2005,taflove_advances_2013}, described by auxiliary differential equations for additional vector fields, such as the current density or polarization fields. While our general method can apply to many such time-domain dispersive models, we consider explicitly here only the Drude model given by
\newline

\begin{equation} \label{Drude}
    \frac{\partial \textbf{J}}{\partial t} + \gamma \textbf{J} -\varepsilon_0 \omega_p^2(t) \textbf{E}=0,
\end{equation}

\noindent where $\textbf{J}$ is the free electron current density inside the medium, $\gamma$ is the Drude damping coefficient, and $\omega_p$ is the plasma frequency. By allowing the plasma-frequency to be time-varying (see, for example, Ref. \cite{baxter_dynamic_nodate}), we are allowing for the material to have a time-varying dispersion.
\newline

For our forward simulation, Eq. \ref{Drude} is solved numerically in the time-domain along with Maxwell’s equations (Eq. \ref{maxwell_general}):

\begin{gather} \label{maxwell2}
 \begin{bmatrix} \varepsilon_0 \varepsilon & 0  & 0\\ 0 & -\mu_0
& 0 \\ 0 & 0 & 1 \end{bmatrix}
\frac{\partial}{\partial t} \begin{bmatrix} \mathbf{E} \\ \mathbf{H} \\ \textbf{J} \end{bmatrix}
+\begin{bmatrix} 0 & -\nabla\times & 1 \\ -\nabla\times & 0 & 0 \\ -\varepsilon_0 \omega_p^2(t) & 0 & \gamma \end{bmatrix}
\begin{bmatrix} \mathbf{E} \\ \mathbf{H} \\ \textbf{J} \end{bmatrix} = \begin{bmatrix} -\mathbf{J}_s(t) \\ 0 \\ 0 \end{bmatrix},
\end{gather}

\noindent The topology is defined not only in $\varepsilon$ (which we take to be static here, but it need not be), but also in the sub-matrices arising from the Drude model (bottom row of the two system matrices), which vanish outside the dispersive medium. In other words, the Drude model is not solved outside the dispersive medium.
\newline

The time-domain dispersive adjoint system equation (from Eq. \ref{General_adjoint}) is

\begin{gather} \label{adjoint2}
 \begin{bmatrix} \varepsilon_0 \varepsilon & 0  & 0\\ 0 & -\mu_0
& 0 \\ 0 & 0 & 1 \end{bmatrix}
\frac{\partial}{\partial t} \begin{bmatrix} \mathbf{E}^{adj} \\ \mathbf{H}^{adj} \\ \textbf{J}^{adj} \end{bmatrix}
+\begin{bmatrix} 0 & -\nabla\times & \omega_p^2(T_m-t)/\omega_{p0}^2 \\ -\nabla\times & 0 & 0 \\ -\varepsilon_0 \omega_{p0}^2 & 0 & \gamma \end{bmatrix}
\begin{bmatrix} \mathbf{E}^{adj} \\ \mathbf{H}^{adj} \\ \textbf{J}^{adj} \end{bmatrix} = \begin{bmatrix} -\mathbf{J}^{adj}_s(t) \\ 0 \\ 0 \end{bmatrix},
\end{gather}

\noindent where, as before, $\textbf{J}_s^{adj}=-\frac{\partial\psi}{\partial\textbf{E}} (T_m-t)$ and where we defined the quantity $\omega_{p0}$ to be a reference plasma frequency, for example, taken at the beginning of the forward simulation $\omega_{p}(t=0)$; note that we used this quantity to define $\textbf{J}^{adj}$ such that it has units of current density.
\newline

For static dispersion, the adjoint problem in Eq. \ref{adjoint2} is very similar to the forward problem in Eq. \ref{maxwell2}; its implications in topology optimization have been recently investigated \cite{hassan_topology_2022}. For time-varying dispersion, the second system matrix of the adjoint problem in Eq. \ref{adjoint2} is fundamentally different than that of the forward problem of Eq. \ref{maxwell2}. As in the previous section, performing the adjoint simulation requires a small modification in the simulation software. The time-varying plasma frequency of the forward simulation must also be time-reversed in the adjoint simulation. 
\newline

Once the forward and adjoint problems are computed, one can use the forward and adjoint fields to compute the gradient of the objective function via

\begin{gather} \label{grad2}
 \frac{\partial F}{\partial p_i}= \int_0^{T_m}\begin{bmatrix} \mathbf{E}^{adj}(T_m-t) \\ \mathbf{H}^{adj}(T_m-t) \\ \textbf{J}^{adj}(T_m - t)/(\varepsilon_0\omega^2_{p0})\end{bmatrix} \cdot \textbf{R}_i(t) dt,
\end{gather}

\noindent where the integrand is the scalar product between the time-reversed adjoint fields and the residue vector, now given by

\begin{gather} \label{residuefull2}
\textbf{R}_i(t) = -\Bigg(\frac{\partial}{\partial p_i} \begin{bmatrix} \varepsilon_0 \varepsilon & 0 & 0\\ 0 & -\mu_0 
 & 0 \\ 0 & 0 & 1 \end{bmatrix}\Bigg)\frac{\partial}{\partial t} \begin{bmatrix} \mathbf{E}(t) \\ \mathbf{H}(t) \\ \textbf{J}(t) \end{bmatrix}
+\Bigg(\frac{\partial}{\partial p_i}\begin{bmatrix} 0 & \nabla\times & -1 \\ \nabla\times & 0 & 0 \\ \varepsilon_0\omega_p^2(t) & 0 & \gamma \end{bmatrix}\Bigg)
\begin{bmatrix} \mathbf{E}(t) \\ \mathbf{H}(t) \\ \mathbf{J}(t) \end{bmatrix} + \frac{\partial}{\partial p_i}\begin{bmatrix} -\mathbf{J}_s(t) \\ 0 \\ 0\end{bmatrix},
\end{gather}

\noindent Again, for pulse shape optimization, only the third term is considered, so the residue is given by Eq. \ref{residue}. This means that, again, no forward fields are required; $\textbf{E}^{adj}$ is required only at the spatial locations where $\textbf{J}_s(t)$ is non-zero.
\newline 

\subsection{Pulse shaping in static materials}\label{staticsection}

In this section we apply our pulse shaping in time method from Section 3 to maximize the transmitted energy through static dielectric and metallic thin films.

\subsubsection{Test case 1: Static, dispersionless thin film}

\begin{figure}
\centering
\includegraphics[width = \linewidth]{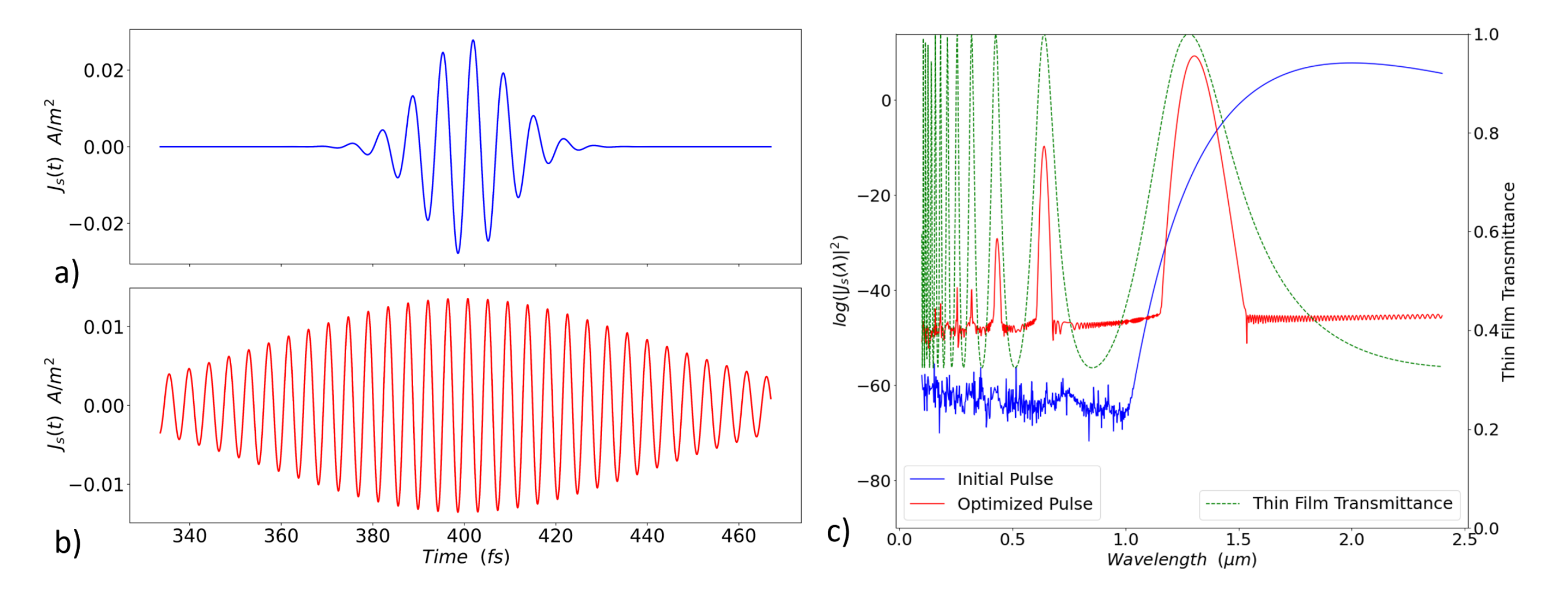}
\caption{a) Initial Gaussian input pulse with duration 13 fs centered at 2 $\mu$m. b) Optimized pulse after 90 iterations. c) Log plot of pulse spectrum for the initial Gaussian pulse (blue line - left vertical axis), and the optimized pulse (red line - left vertical axis), overlayed with the transmittance spectrum of the dielectric thin film (green dashed line - right vertical axis).}
\label{static_perm}
\end{figure}

\begin{figure}
\centering
\includegraphics[width = 0.5\linewidth]{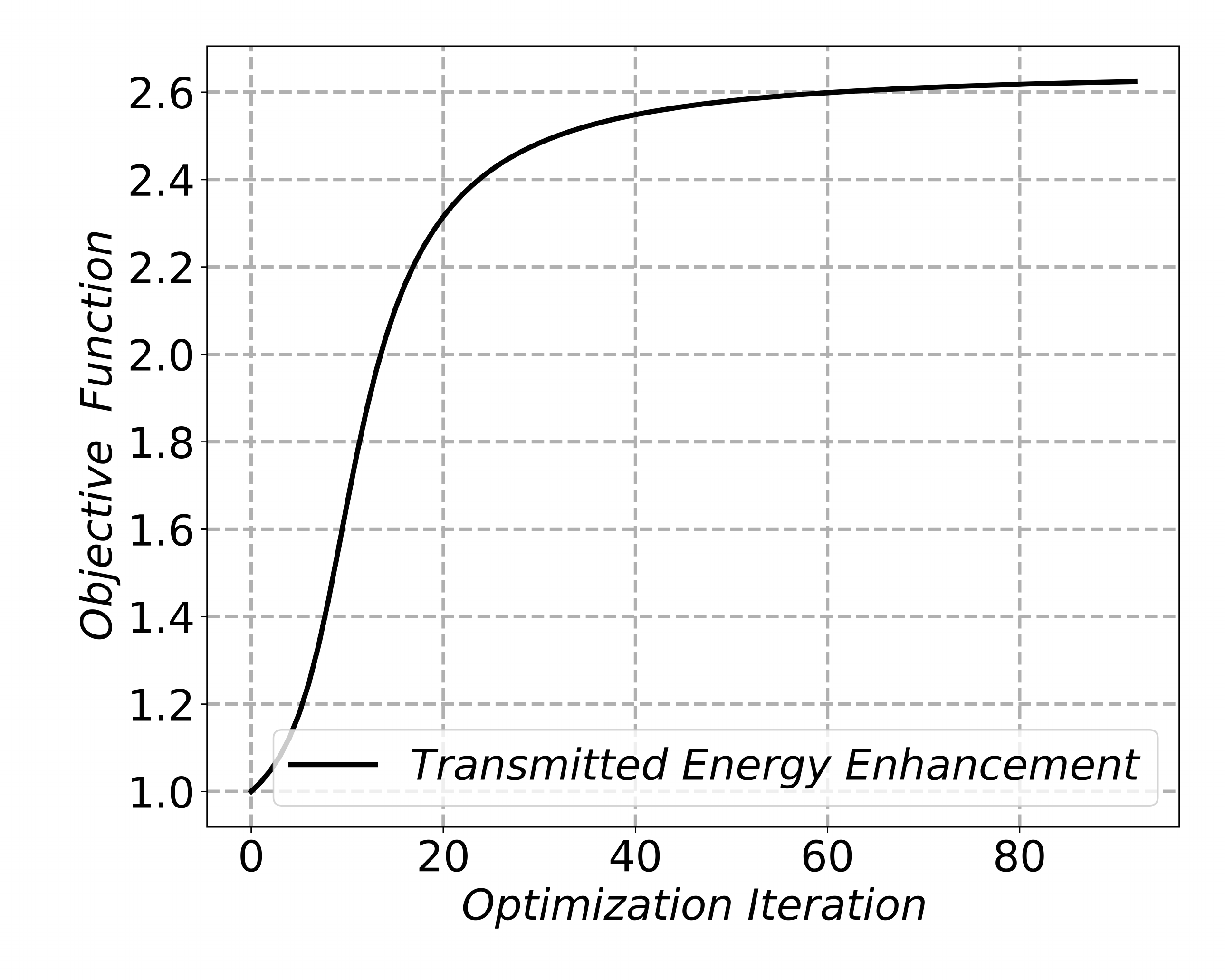}
\caption{Objective function evolution during optimization procedure for a static, dispersionless thin film.}
\label{static_perm_FOM}
\end{figure}

As a test case, we consider a 200 nm dielectric thin film with static permittivity $\varepsilon=10.25$. As plotted in Fig. \ref{static_perm}a, our initial pulse is a Gaussian with a central wavelength of $\lambda_0=2$ $\mu$m and pulse duration $\tau=13$ fs. The center wavelength is purposefully chosen to be in a spectral range of low transmittance through the dielectric film. The evolution of the objective function during the optimization is plotted in Fig. \ref{static_perm_FOM}, showing a 2.6$\times$ enhancement in transmitted energy after only 90 iterations of our pulse shaping algorithm.
\newline

The optimized pulse is plotted in Fig. \ref{static_perm}b, and has a much larger pulse width than the initial pulse (only a portion of the pulse is shown here, in the same time window as the initial pulse in  Fig. \ref{static_perm}a). In Fig. \ref{static_perm}c, the spectrum of the initial pulse (solid blue line), and optimized pulse (solid red line) are plotted corresponding to the left-vertical axis, with the transmittance spectrum of the dielectric thin film (dashed green line) corresponding to the right-vertical axis. Here we can see that, indeed, the spectrum of the initial pulse is in a bandwidth of low transmittance. The pulse shaping algorithm added frequency components centered at the Fabry-Perot resonances of the thin film, thus achieving the goal of increasing the transmitted energy without increasing the pulse energy. The large width of the optimized pulse is due to the optimizer keeping the bandwidth narrow within each resonance to ensure a higher transmission.

\subsubsection{Test case 2: Static, Drude-metal thin film}

\begin{figure}
\centering
\includegraphics[width = \linewidth]{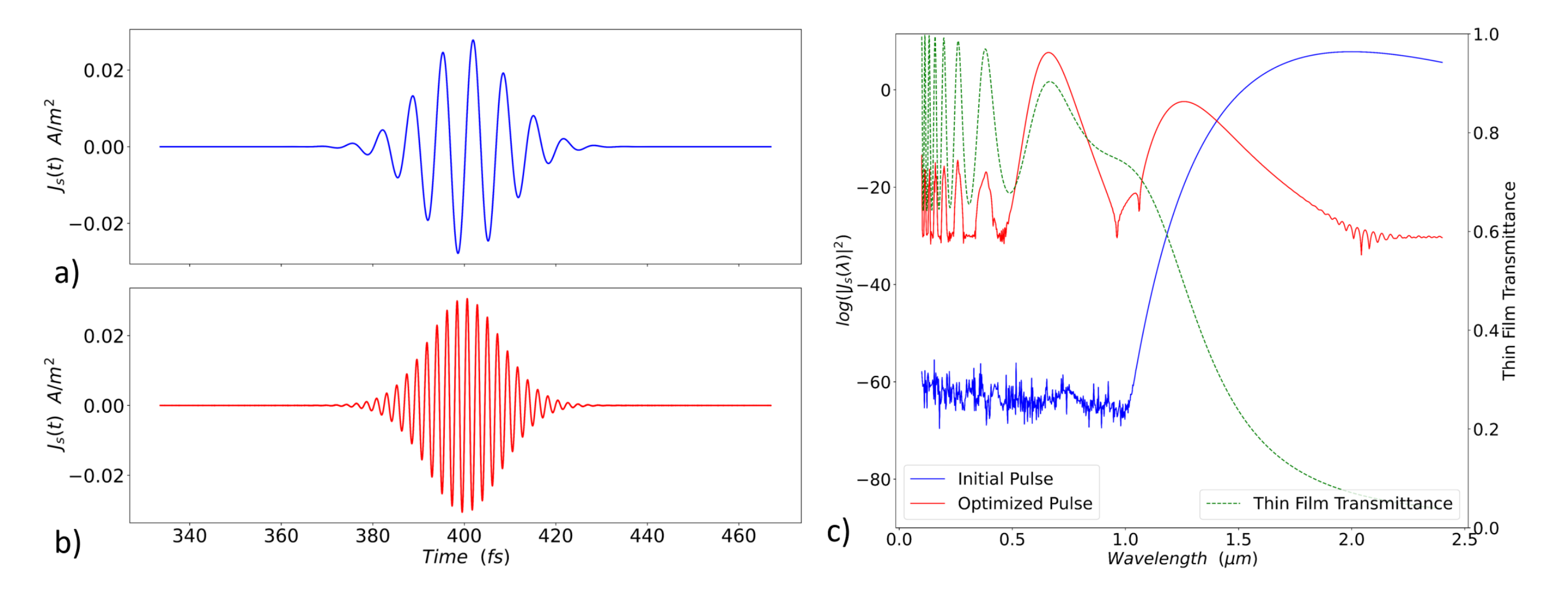}
\caption{a) Initial Gaussian input pulse with duration 13 fs centered at 2 $\mu$m. b) Optimized pulse after 40 iterations. c) Log plot pulse spectrum of the initial Gaussian (blue line - left vertical axis), the optimized pulse (red line - right vertical axis) overlayed with the transmittance spectrum of the metallic thin film (green dashed line - right vertical axis).}
\label{static_drude}
\end{figure}

As a second test case, we will demonstrate pulse shapingin time with a 200nm metallic thin film with an optical response described by the Drude model (Eq. \ref{Drude}),  where $\omega_p=3\times 10^{15}$ rad/s is the static plasma frequency, $\gamma=2\times 10^{14}$ rad/s, and with a background permittivity of $\varepsilon=4$. As plotted in Fig. \ref{static_drude}a, our initial pulse is a Gaussian with a central wavelength of $\lambda_0=2$ $\mu$m and duration $\tau=13$ fs. Once again, the center wavelength is purposefully chosen to be in a spectral range of low transmittance through the metal film. After 40 iterations of pulse shaping in time, we achieve a transmitted energy that is 12$\times$ higher than for the initial pulse. The optimized pulse is plotted in Fig. \ref{static_drude}b. 
\newline

Fig. \ref{static_drude}c shows a log plot of the frequency spectra of the initial pulse (solid blue line) and the optimized pulse (solid red line), which correspond to the left-vertical axis. The green, dashed line (right axis) is the transmittance of the thin, metallic film. We again see that the new pulse has gained frequency components at the Fabry-Perot resonances of the thin film.

\newpage

\begin{backmatter}
\bmsection{Funding}
place holder text 

% Content in the funding section will be generated entirely from details submitted to Prism. Authors may add placeholder text in the manuscript to assess length, but any text added to this section in the manuscript will be replaced during production and will display official funder names along with any grant numbers provided. If additional details about a funder are required, they may be added to the Acknowledgments, even if this duplicates information in the funding section. See the example below in Acknowledgements.

\bmsection{Acknowledgments}
The authors would like to thank Prof. Antonio Cal\`{a} Lesina, Prof. Israel De Le\'{o}n, and Dr. Orad Reshef for helpful discussions.

\bmsection{Disclosures}
The authors declare no conflicts of interest.

\bmsection{Data Availability}
Data underlying the results presented in this paper can be generated via the python scripts in Supplement 1.

\bmsection{Supplemental document}
See Supplement 1 for supporting content. 
\end{backmatter}

%%%%%%%%%% If using BibTeX:
\bibliography{Time-Adjoint-Paper}

\end{document}